# Exploring possibilities of band gap measurement with off-axis EELS in TEM

Svetlana Korneychuk[1], Bart Partoens[2], Giulio Guzzinati[1], Rajesh Ramaneti[3,4], Joff Derluyn[5], Ken Haenen[3,4] and Jo Verbeeck[1]

[1]*Electron Microscopy for Material Science (EMAT), University of Antwerp, 2020 Antwerp, Belgium*
[2]*Condensed Matter Theory (CMT), University of Antwerp, 2020 Antwerp, Belgium*
[3]*Institute for Materials Research (IMO), Hasselt University, 3590 Diepenbeek, Belgium.*
[4]*IMOMEC, IMEC vzw, 3590, Diepenbeek, Belgium.*
[5]*EpiGaN NV, 3500, Hasselt, Belgium.*

**Abstract**

A technique to measure the band gap of dielectric materials with high refractive index by means of energy electron loss spectroscopy (EELS) is presented. The technique relies on the use of a circular (Bessel) aperture and suppresses Cherenkov losses and surface-guided light modes by enforcing a momentum transfer selection. The technique also strongly suppresses the elastic zero loss peak, making the acquisition, interpretation and signal to noise ratio of low loss spectra considerably better, especially for excitations in the first few eV of the EELS spectrum. Simulations of the low loss inelastic electron scattering probabilities demonstrate the beneficial influence of the Bessel aperture in this setup even for high accelerating voltages. The importance of selecting the optimal experimental convergence and collection angles is highlighted. The effect of the created off-axis acquisition conditions on the selection of the transitions from valence to conduction bands is discussed in detail on a simplified isotropic two band model. This opens the opportunity for deliberately selecting certain transitions by carefully tuning the microscope parameters.
The suggested approach is experimentally demonstrated and provides good signal to noise ratio and interpretable band gap signals on reference samples of diamond, GaN and AlN while offering spatial resolution in the nm range.

1. Introduction

Electron energy loss spectroscopy (EELS) in transmission electron microscopy (TEM) provides information about the local dielectric properties of materials including band gap information in semiconducting materials. EELS has an advantage over conventional techniques for band gap measurement by providing much better spatial resolution and the possibility to map the band gap values, for example, in a multilayered heterostructure [1–4].

So far however, low loss EELS is not widely used for band gap measurements due to the presence of "parasitic" unwanted losses which are usually superimposed to the band gap signal. These unwanted losses occur e.g. through retardation effects via the emission of Cherenkov radiation [5,6] and the excitation of surface-guided light modes. Cherenkov radiation occurs when a charged particle (such as an electron) travels through a dielectric medium with a speed greater than the speed of light in this medium [7]. The threshold speed for Cherenkov transitions is determined by the ratio $v > c/n$, where $v$ – is the speed of a charged particle in the material and $c$ – the speed of light in vacuum. For materials with high refractive index, Cherenkov radiation in the TEM is in many cases unavoidable. The challenges of measuring the band gap in the TEM are well understood and have been discussed in great detail [8–13] with many propositions on how to overcome them [14–18].

Fortunately, according to the simulations based on the so-called Kröger formula [19] that describes low loss inelastic scattering in a plan parallel sample geometry, all the unwanted losses happen at very low scattering angles and can, in principle, be avoided by not recording scattering events in that scattering range [9,20]. In this work we explore how a conical illumination scheme allows to eliminate the undesired signals and to accurately measure the bandgap in high-refractive-index semiconductors, while retaining a high spatial resolution, the key characteristic of the TEM. We will experimentally demonstrate that the proposed setup indeed provides excellent suppression of the unwanted loss signals at both high (300 keV) and low (60 keV) beam energies while at the same time strongly suppressing the zero loss peak. This suppression drastically simplifies the retrieval of the band gap signal and significantly improves the signal to noise ratio.

Besides spatial resolution, there is another important difference setting EELS apart from optical techniques. While for photons the momentum transfer to the material making a transition from the valence to conduction band is approximately equal to zero and negligible, in EELS the transfer of momentum between the fast electron and the sample can be substantial [21].

This provides the possibility in EELS to obtain information about all possible interband transitions with the attractive opportunity to deliberately select certain transitions while excluding others. The role of this momentum transfer selection in the proposed setup will be clarified and prospects for applications are outlined.

2. **Experimental set-up**

The unwanted Cherenkov and surface losses are characterized by only a small momentum transfer, as becomes obvious when numerically evaluating the Kröger formula [8]. Indeed, these unwanted inelastic interactions produce scattering angles in the μrad range (see. fig.1 (b,c)), while Bragg scattering typically occurs in the order of mrad for electrons in the 100's keV range. This opens the possibility of avoiding the unwanted part of inelastic scattering by placing the spectrometer entrance aperture off-axis with respect to the unscattered electron beam in the diffraction plane. To first approximation, we can model the intensity distribution in this diffraction plane as a convolution of (i) the inelastic scattering cross section, assuming a homogeneous medium with a given dielectric function, (ii) the elastic scattering contribution consisting of Bragg peaks due to the atomic structure of the material and (iii) the angular distribution of the incoming electron beam. For plane wave illumination, this results in a series of Bragg spots that are broadened by inelastic scattering. This broadening now contains a small contribution from Cherenkov radiation and a larger one from inelastic scattering, therefore off-axis spectral recording can easily avoid the lowest angles thus suppressing the unwanted Cherenkov effect [12,22,23]. With convergent beam illumination, we obtain better spatial localization, as the probe is now focused on the sample, but lose momentum resolution as different scattering features are now becoming harder to disentangle especially if also the convergent beam electron diffraction (CBED) discs start to overlap and off-axis spectral detection becomes problematic.

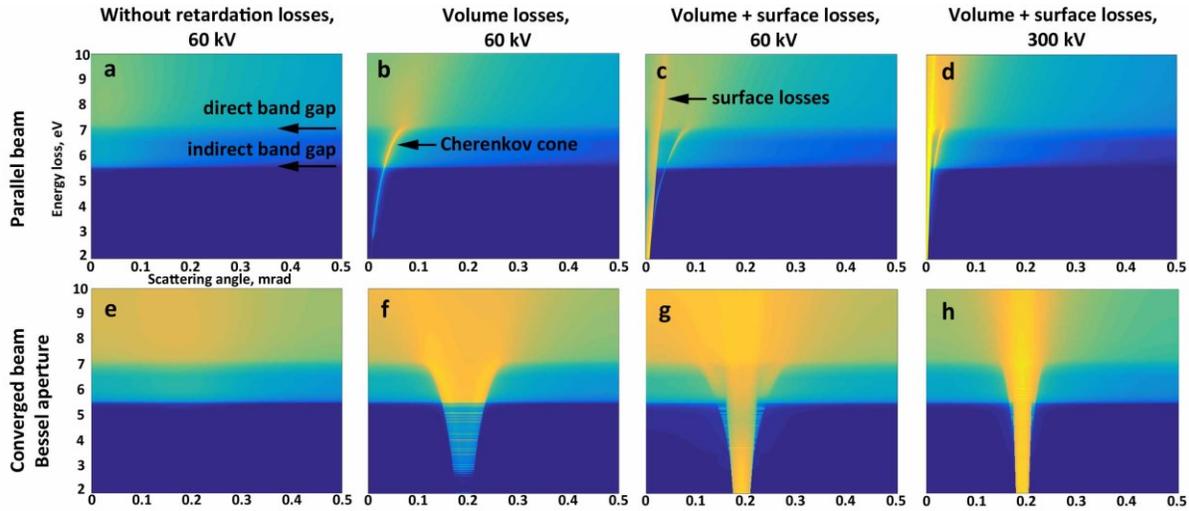

*Fig.1. Simulated double differential inelastic cross section for a plan parallel diamond thin film (50 nm) at a beam energy of 60 keV and 300 keV as a function of energy loss E and scattering angle θ based on the Kröger equation [24]. Illumination with either a parallel beam or a Bessel beam (0.2 mrad opening angle) is simulated including either only the non-relativistic loss function Im(1/ε) (a, e), including relativistic effects (b, f) and including surface effect (c, d, g, h). Note the presence of clear Cherenkov and surface losses at scattering angles only below 0.1 mrad and even lower for 300 keV electrons. Using the Bessel beam, shifts these unwanted losses to the opening angle of the beam and creates a region around the center which is free of these unwanted excitations. Stripes at the simulation are caused by numerical artifacts due to noise in the tabulated dielectric function.*

A solution can be found by using an annular aperture (fig.2a) in the condenser plane of the microscope. This kind of aperture creates an approximation to an electron Bessel beam [25–28] and is accordingly called a Bessel aperture. This aperture creates ring-shaped CBED pattern due to the conjugation of condenser and diffraction planes in STEM mode, resulting in an assembly of rings in the diffraction plane as sketched in Figure 2.

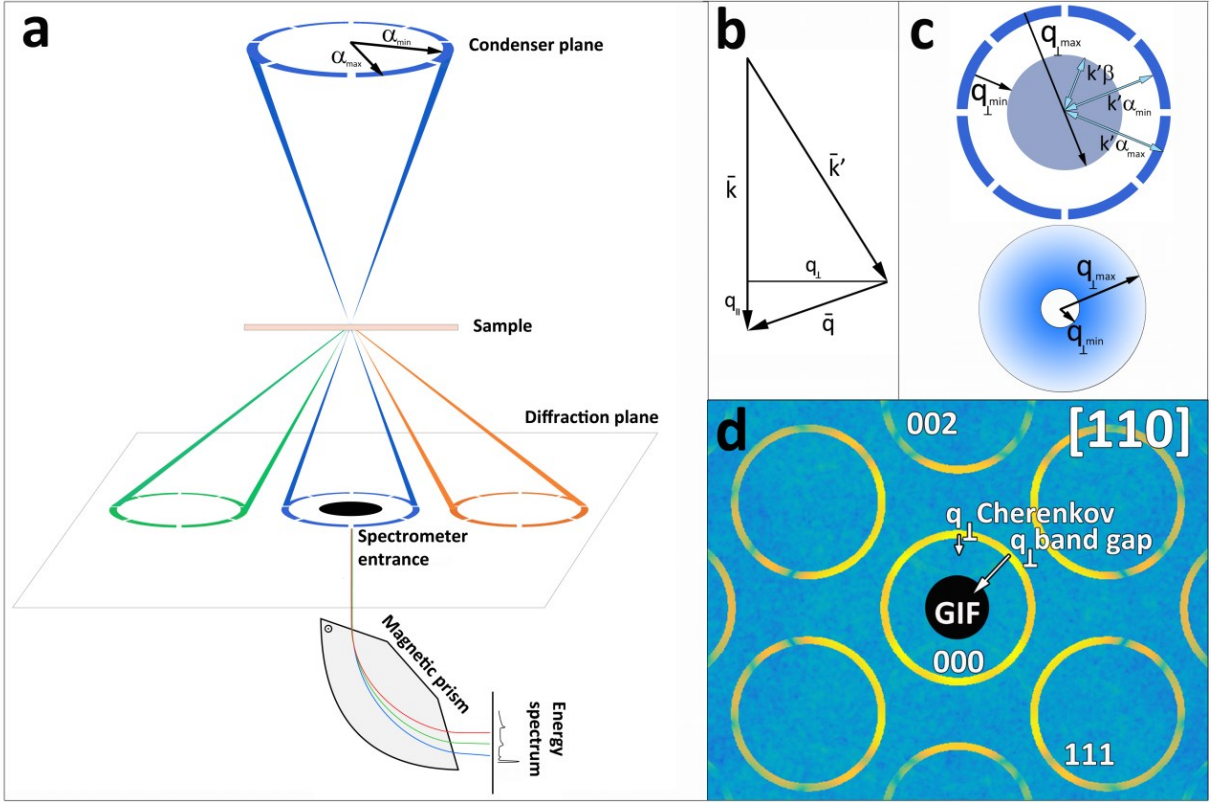

*Fig.2. a. Schematic diagram of the off-axis experiment with a Bessel aperture. b. Vector plot of the transfer of momentum in inelastic electron scattering. c. Limits imposed to this transfer of momentum when placing the entrance aperture (central disc) of the spectrometer in the center of the Bessel aperture. d. Simulated CBED pattern [29] of diamond [110] with a Bessel aperture at 60 kV acceleration voltage and 10 mrad convergence angle showing the diffracted rings and the position of the spectrometer entrance aperture. Note how small momentum transfers stemming from Cherenkov radiation can't bring electrons into the spectrometer while larger momentum transfers originating from conventional inelastic losses do end up in the spectrometer. Note also that elastic scattering does not enter the spectrometer and therefore no zero loss peak is expected in the spectrum.*

By selecting the appropriate camera length (CL) and convergence angle, the central (000) reflection can be made to surround the spectrometer entrance aperture, thus selecting only these scattering events within a certain range of momentum transfers (fig.2d).

It is important to highlight that using this Bessel aperture, the scattering distribution will be altered in a fundamental way and can be described by the convolution of the double differential inelastic scattering cross section with the Bessel aperture function. Figure 1 e-h demonstrates the result where now the unwanted losses inside the band gap region of the material are shifted to the angles of the ring aperture. This area now contains a significant amount of retardation losses, and emerging from it, we can see the Cherenkov cone. The area in the center of the aperture appears free from Cherenkov losses and still contains information of interest on the band gap of the material as long as the opening angle of the ring remains very small with respect to q-dependent changes in the band structure. The figure has been computed for small angles in order to resolve the retardation losses better, while in practice slightly higher angles will be preferred due to the corresponding higher spatial resolution and convenience of staying in the same operating mode in our microscope set-up.

The Bessel aperture can provide an advantage over a standard round one for low loss STEM-EELS experiments by circularly averaging over the momentum transfer thus avoiding complications with possible anisotropic features [15].

A major advantage of the setup is the rejection of both the elastically scattered beam as well as the low angle Cherenkov and surface scattering. Indeed, removing the so-called zero loss peak from the spectrum has several advantages as the full dynamic range of the spectrometer can be used to record the loss signal, no subtraction of the zero loss peak is needed and the signal to noise ratio is greatly improved as the undesired (tail of the) elastic signal is suppressed, reducing background and the associated Poisson noise.

Therefore, to simplify the setup the convergence angle of the Bessel beam should be chosen smaller than the Bragg angles in the material in order to keep the rings from overlapping. The situation where they almost touch is considered ideal as it provides maximum suppression of both Cherenkov losses and elastic signal while preserving good spatial resolution, together with a collection angle chosen to be just smaller than a convergence angle. However, working at these conditions will inevitably blur some of the q-dependent band structure features which may or may not be desired, but for q-averaged band gap mapping this could be considered ideal. While the experiment can still be performed in case of overlapping rings, the proper positioning of the aperture between the various reflections and the interpretation becomes increasingly difficult together with reducing of circular averaging advantage.

On the other hand working at smaller opening angles will compromise the spatial resolution of the setup due to the diffraction limit of the probe size [30]. The non-overlap criterion will prevent true atomic resolution with this setup, but in the low loss regime, the spatial resolution is in any case likely dominated by delocalization of the inelastic scattering [21]. The advantage of this regime is that q-dependent transitions can be more carefully selected without the blurring effect that occurs for larger opening angles.

3. q selection

Unlike the optical measurements the transferred momentum $q$ of the transitions from valence to conduction band can be significantly different from zero in EELS experiments while the primary energy of the electron beam is more than enough to excite all possible interband transitions. This leads to a typical low loss spectrum which is very hard to disentangle into a meaningful band structure picture.

Working off-axis with the proposed Bessel setup on the other hand allows to post-select a small subset of the possible transitions by enforcing strict momentum transfer selection criteria, which could make direct interpretation of the low loss EELS spectrum more manageable in terms of band structure. This post-selection acts only on the perpendicular component $q_\perp$ of the momentum transfer (fig. 2b) where the upper limit $q_{\perp max}$ of the $q_\perp$ interval is determined by the sum of the maximum angle $\alpha_{max}$ of the Bessel aperture and the maximum spectrometer entrance aperture angle $\beta$ as $q_{\perp max} = k'\sin\theta \approx k'\theta = k_0\sqrt{1-2\theta_E}(\alpha_{max} + \beta)$, where forward momentum $k_0 = \frac{2\pi}{\lambda}$ and characteristic scattering angle $\theta_E \approx \frac{E}{2E_0}$ neglecting relativistic corrections [11]. The lower limit $q_{\perp min}$ is given by the difference of the lowest angle in the Bessel aperture with the spectrometer entrance aperture as $q_{\perp min} = k_0\sqrt{1-2\theta_E}(\alpha_{min} - \beta)$ (fig. 2c). The relation between $q$, $\theta$ and $\theta_E$ is given by:

$$q^2 = q_{\|}^2 + q_{\perp}^2 = k_0(1 - \sqrt{1 - 2\theta_E} \cos\theta + 1 - 2\theta_E) \approx k_0^2(\theta^2 + \theta_E^2) \qquad (1)$$

More detailed description of inelastic scattering vector geometry is outlined in this work [31].

Choosing an appropriate interval via tuning the Bessel aperture geometry with respect to the spectrometer collection aperture allows to select only those transitions that are of interest. The allowed $q_{\perp}$ vectors effectively form a donut with radius $α$ and width $β$ assuming the Bessel ring is very narrow compared to the entrance aperture. This ring effectively averages out anisotropic variations in the band structure which may or may not be an advantage depending on what information is desired. The transitions in the q-donut with lower $q_{\perp}$ will have higher intensity due to purely geometrical reasons – more of them can end up in the spectrometer entrance than transitions with higher $q_{\perp}$.

In order to understand the effect of this $q$ selection on the low loss spectra, we have to compute the joint density of states (JDOS) and then we need to add the typical EELS cross section considerations for low loss excitations.

In order to shed light on this rather complicated combination of factors, we demonstrate the effect on a toy model band structure consisting only of two parabolic bands (fig. 3a), following arguments laid out for *q=0* transitions in [32,33].

Let's assume that a transition from the valence to conduction band occurs with initial wave vector *k*$_i$ and final wave vector *k*$_f$. The conduction band dispersion relation is given as:

$$E_{cb} = E_c + \frac{\hbar^2 k_f^2}{2m_e} \qquad (2)$$

where $E_c$ is the bottom of the conduction band energy and $m_e$ is the conduction electron's effective mass. For the valence band the dispersion relation is:

$$E_{vb} = E_v - \frac{\hbar^2 k_i^2}{2m_h} \qquad (3)$$

where $E_v$ is the top of the valence band energy and $m_h$ is the hole's effective mass.

The energy transferred from the fast electron to the sample (energy loss) will then be:

$$E = E_{cb} - E_{vb} = E_{bg} + \frac{\hbar^2}{2}\left(\frac{k_i^2}{m_h} + \frac{k_f^2}{m_e}\right) \qquad (4)$$

where $E_{bg}$ denotes the band gap energy.

Applying momentum conservation we get:

$$\hbar \boldsymbol{k}_i + \hbar \boldsymbol{q} = \hbar \boldsymbol{k}_f \qquad (5)$$

or:

$$\boldsymbol{k}_f = \boldsymbol{k}_i + \boldsymbol{q} \qquad (6)$$

Then the expression (3) for energy loss can be rewritten as:

$$\frac{\hbar^2}{2}\left(\frac{k_i^2}{m_h} + \frac{(\mathbf{k}_i + \mathbf{q})^2}{m_e}\right) + E_{bg} - E = 0 \tag{7}$$

The number of transitions with this energy E is now given by the joint density of states (JDOS):

$$JDOS(E, \mathbf{q}) = 2 \int_{BZ} \frac{d\mathbf{k}}{(2\pi)^3} \delta\left(\frac{\hbar^2}{2}\left(\frac{k_i^2}{m_h} + \frac{(\mathbf{k}_i + \mathbf{q})^2}{m_e}\right) + E_{bg} - E\right) \tag{8}$$

To calculate this integral numerically, a Lorentzian representation of the δ-function can be used:

$$\delta(x) = \lim_{\epsilon \to 0} \frac{\epsilon/\pi}{x^2 + \epsilon^2}$$

To test the numerical integration, the $\mathbf{q} \sim 0$ case can be considered. For this case, the JDOS can also be calculated analytically as follows (with $m_r = m_e m_h / m_e + m_h$):

$$JDOS(E, 0) = 2 \int_{BZ} \frac{d\mathbf{k}}{(2\pi)^3} \delta\left(\frac{\hbar^2 k^2}{2m_r} + E_{bg} - E\right) \tag{9}$$

$$= \frac{8\pi}{(2\pi)^3} \int k^2 dk\, \delta\left((k - \sqrt{\frac{2m_r}{\hbar^2}(E - E_{bg})})(k + \sqrt{\frac{2m_r}{\hbar^2}(E - E_{bg})}) \frac{\hbar^2}{2m_r}\right) \tag{10}$$

$$= \frac{1}{\pi^2} \frac{2m_r}{\hbar^2} \frac{1}{2\sqrt{\frac{2m_r}{\hbar^2}(E - E_{bg})}} \int k^2 dk\, \delta\left((k - \sqrt{\frac{2m_r}{\hbar^2}(E - E_{bg})})\right) \tag{11}$$

$$= \frac{1}{2\pi^2}\left(\frac{2m_r}{\hbar^2}\right)^{3/2} \sqrt{E - E_{bg}} \tag{12}$$

The square root dependence of JDOS on energy is also given in literature [34,35].

We now find the EELS double differential scattering cross section in dipole approximation [36,37]:

$$\frac{\partial^2 \sigma}{\partial E \partial \Omega} \propto \frac{1}{k_0^2(\theta^2 + \theta_E^2)} JDOS(q, E), \tag{13}$$

Taking into account the selected q-donut (fig.3) the total cross-section is given by:

$$\frac{d\sigma}{dE} = \int d\Omega \frac{\partial^2 \sigma(\theta)}{\partial E \partial \Omega} \propto \int_{\alpha_{min} - \beta}^{\alpha_{max} + \beta} \sin\theta d\theta \int_0^{2\pi} d\phi \frac{1}{k_0^2(\theta^2 + \theta_E^2)} JDOS(E, q), \tag{14}$$

These cross sections are shown in Figure 3d, for the cases $[\alpha_{min} - \beta, \alpha_{max} + \beta]$ = [0 mrad, 1 mrad], [1 mrad, 2 mrad] and [5 mrad, 10 mrad]. These simulations demonstrate that band gap value simulated for the on-axis case (0-1 mrad) matches the value obtained slightly off-axis (1-2 mrad), meanwhile the signal obtained at larger angles significantly shifts the band gap towards higher energies. This suggests that the best strategy for band gap measurements is to keep the distance between spectrometer entrance and Bessel aperture (basically, $\mathbf{q}_{\perp min}$) as small as possible though still avoiding the retardation losses. Also it's important to highlight that the smaller convergence angle is the more prominent the direct band gap onset will be due to the higher momentum resolution.

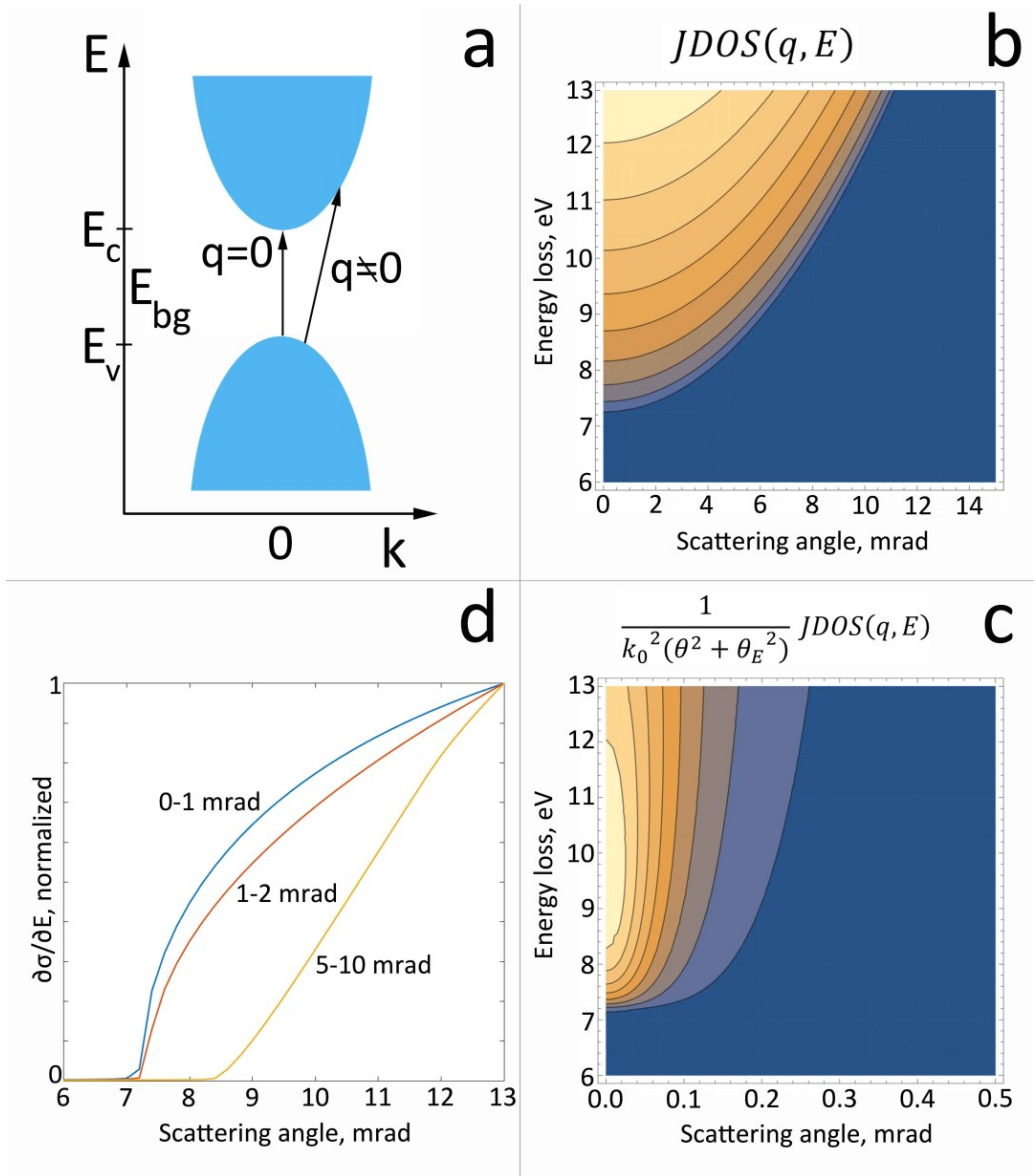

*Fig.3. a. A schematic two band model for excitations in a material with two parabolic bands. b. JDOS dependence on the momentum transfer q (shown through the scattering angle $\theta \approx \frac{q}{k_0}$) calculated for the model. Value of $E_{bg}$ was taken as 7.2 eV in accordance of direct bad gap of diamond. Effective masses similar to diamond were taken as $m_e=0.57m_0$ amd $m_h=0.8m_0$ [38]. c. Simulation of the double differential EELS cross-section for all q. The Lorentzian distribution of inelastic scattering around low scattering angles is clearly visible and selects predominantly transitions close to q=0. d. Simulated EELS spectra extracted from (c) for different choices of q calculated using expression (13). Note that the onsets will shift towards higher values when increasing the selected q. This constitutes one more reason to use small convergence angles and stay well in the first Brillouin zone.*

## 4. Results

The proposed method was tested on a multilayered semiconductor heterostructure containing nanocrystalline diamond (NCD), AlGaN, GaN and AlN layers. Two alternative conditions were applied to probe the effectiveness of the technique – using an acceleration voltage of 60 keV which should allow to minimize the Cherenkov losses and using 300 keV where the unwanted effects are much more pronounced. GaN, AlN and AlGaN being direct semiconductors with rather flat band structure are good objects to compare the band gap values estimated by EELS with the standard ones obtained by optical techniques that only 'see' the direct band gap.

Off-axis EELS measurements as proposed here can be influenced by several experimental details that need to be carefully considered. Starting out with an experiment without a sample already introduces a few peculiarities. As can be seen from figure 4a, the intensity of the zero loss peak (ZLP) acquired in off-axis conditions with Bessel aperture is suppressed by more than 5 orders of magnitude as compared to the same experimental conditions when replacing the Bessel aperture with a standard round aperture. Nevertheless, a ZLP is still weakly present in the spectra while this was not expected.

Somehow, a fraction of intensity ends up being elastically scattered to the center of the ring in the diffraction plane which could be due to: inelastic phonon scattering, decoherence effects [39], transmission of electrons through the opaque region of the Bessel aperture or less than ideal focusing of the Bessel aperture in the diffraction plane. It turns out, that the dominant effect here is caused by a slight misalignment of the height of the Bessel condenser aperture which leads to a slightly defocused elastic image in the diffraction plane as shown in fig.4c. A typical Arago or Poisson spot becomes visible in the center which brings elastic electrons back on the optical axis due to constructive interference effects [30]. This effect is particularly strong for smaller convergence angles. Ideally this effect would be corrected by mechanically lifting the aperture to a different position, but here we have to compensate this effect by slightly defocusing the diffraction lens.

Another artefact comes from inelastic scattering to the surface plasmon in the 1 μm thick Au film that is used to fabricate the Bessel aperture. This results in a peak at 2.5 eV that is visible in fig 4.a.

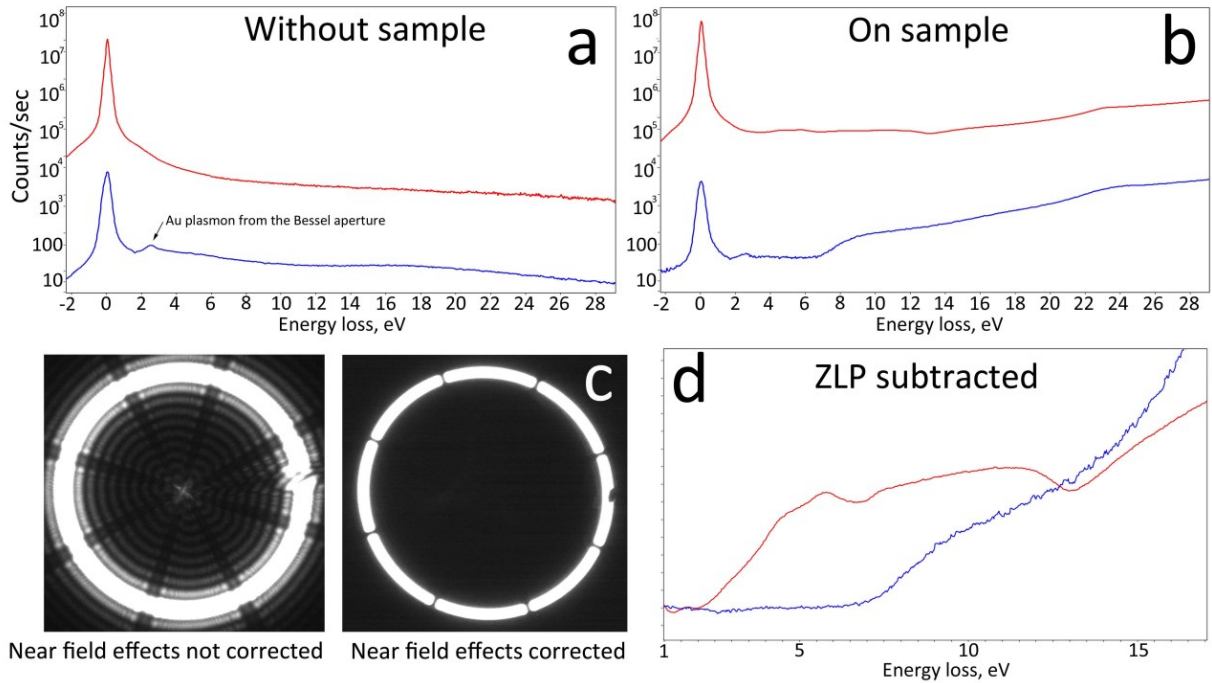

*Fig.4. EELS spectra taken at 300 kV, 1.5 mrad convergence angle, 1s exposure in off-axis configuration with Bessel aperture (blue) and on-axis configuration with standard round aperture (red) a. without a sample. b. on nanocrystalline diamond. c. Diffraction pattern without sample without refocusing of the diffraction lens showing an Arago spot to appear in the center (left) and with corrected diffraction plane height (right) d. Background removed spectra from (b) by subtraction of the spectrum without a sample in (a). All the spectra are corrected for the same electron dose by taking into account the difference between the areas of the Bessel aperture and conventional round aperture.*

Using the same experimental conditions on a polycrystalline diamond film resulted in spectra presented in fig 4b. The 1.5 mrad convergence angle avoids the intersection of Bessel rings due to Bragg reflection occurring at ~10 mrad for diamond at 300 keV.

Comparing the intensity of the ZLP for the Bessel case with and without sample results in a very small reduction of the intensity due to absorption and scattering while quasi elastic phonon scattering [40] and other very low loss excitations apparently do not contribute significantly to this peak when compared to the case for the conventional round aperture.

In conclusion we observed that the dominant contribution is likely coming from a residual slight defocus of the diffraction plane, but the 4$^{th}$ order of magnitude suppression of the ZLP is already a very welcome effect to significantly increase the reliability of the background removal step and to lower the dynamic range of the spectrum for recording on a detector.

Background subtracting the low loss spectra with the no-sample spectra (fig. 4d) shows a significant reduction of unwanted losses for the Bessel configuration even at 300 keV where the Cherenkov losses are typically very pronounced for diamond (fig. 1 d,h). This experiment shows the effectiveness of the proposed technique even at high acceleration voltages and for high refractive index materials.

It's worth noting that even at 60 keV which is, sometimes, considered as a 'safe' acceleration voltage for band gap measurements, Cherenkov losses in high refractive index materials like diamond (n=2.4) cannot be completely eliminated (fig.1).

Diamond, being an indirect semiconductor represents a complication due to the presence of two band gaps – an indirect one at ~5.5 eV and a direct one at ~7.4 eV [38]. Simulated double differential inelastic cross section based on the tabulated dielectric function of diamond [41] clearly shows both band gap onsets (fig.1). In the experimental spectra we can clearly identify only the direct band gap at 7.4 eV due to, the dominance of the scattering angles close to $q$=0 via the Lorentzian distribution of the inelastic scattering. This makes the result similar as for an optical measurement at $q$=0 for direct transitions. The indirect band gaps may be measured by selecting $q$ in the different range. This attractive possibility will be investigated in future work, and here, we concentrate on eliminating the retardation losses and improving the direct band gap signal.

In order to demonstrate the main advantage of STEM-EELS for band gap measurements – spatial resolution– we obtained EELS line profiles over multilayered lamella (fig.5a) at 60 keV. A slightly higher convergence angle of 8 mrad is used in order to increase the spatial resolution while still avoiding Bragg reflections to overlap. The results are presented in fig.5.

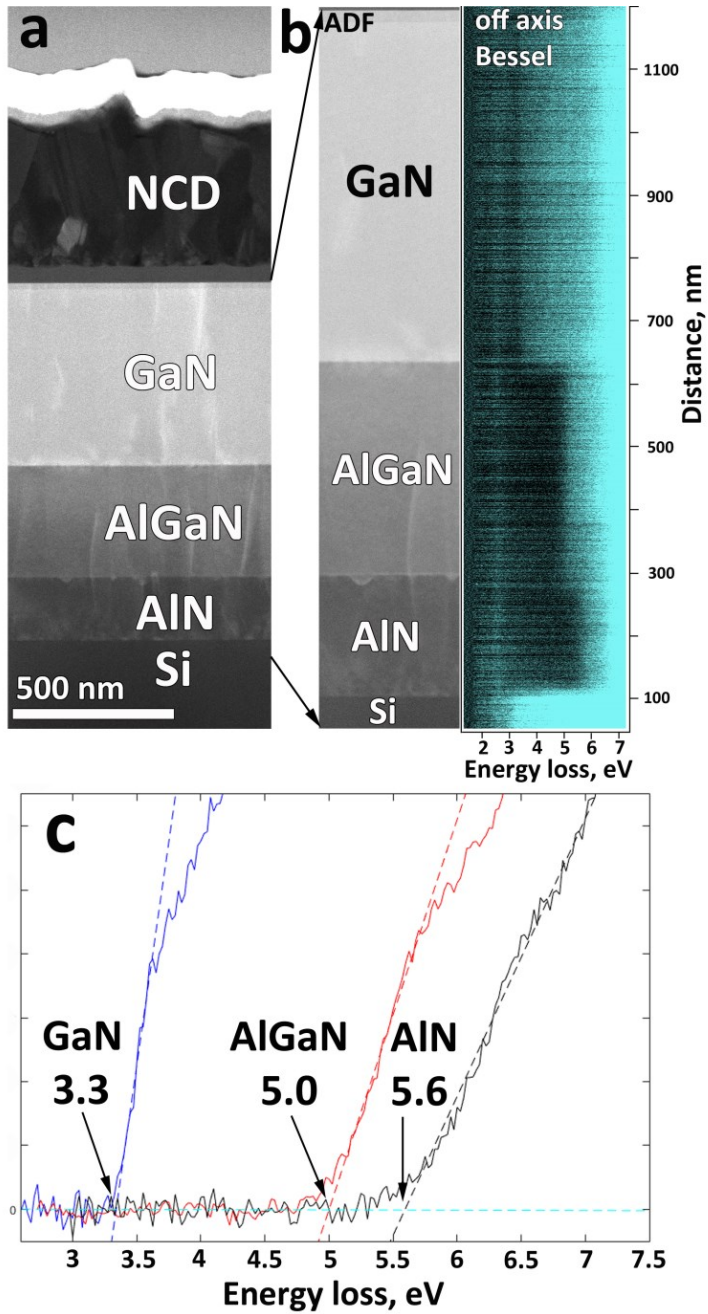

*Fig.5. a. ADF-STEM image of a FIB lamella prepared from multilayered heterostucture NCD-GaN-AlGaN-AlN. b. Background subtracted EELS line profile along the GaN-AlGaN-AlN layers performed in Bessel configurations at 60 keV and opening angle 8 mrad showing good spatial resolution. c. EELS low loss spectra extracted from the line profile (b) demonstrating band gap onsets for GaN, AlGaN and AlN layers with approximate fitting of band gap values.*

The proposed configuration gives an advantage in signal to noise ratio (SNR) when acquiring low loss spectra as the ZLP intensity is strongly suppressed making higher exposure times possible without saturating the detector. The same SNR can be achieved on axis only when adding up a significant amount of spectra with lower exposure times.

The background subtracted low loss spectra acquired when scanning over the epitaxial GaN, AlGaN and AlN layers demonstrates the mapping of clear band gap onsets at relatively high spatial resolution. As these materials are direct semiconductors with relatively flat band structures, the results for GaN are in good agreement with optical methods [42]. Lower band gap value for AlN and higher value for AlGaN can be the sign of Ga contamination for AlN and Ga depletion for AlGaN introduced during the synthesis. It has to be noted however that optical methods could never obtain the spatial resolution presented here which is estimated to be of the order of a few nm.

Reaching the polycrystalline diamond layer, however, the situation is more complicated and the resulting spectra are displayed in fig.6.a. Note that not in all grains of the polycrystalline sample a clear band gap signal is obtained. The reason for this is linked to different grain orientations leading to a complicated appearance of Bragg spots, causing changes in the length of $q_{\perp min}$ and $q_{\perp max}$ together with the shape of q-selection and, therefore, introducing directional effects and collecting the transitions with much higher values of $q$ than for the direct band transitions. This can lead, firstly, to significant decrease of band gap signal and, secondly, to the shift of the band gap onset to higher values.

This effect can be overcome by choosing a smaller convergence angle of 0.5 mrad (fig 6.b) but unfortunately in our current setup, the correction of the diffraction plane position leads to problems with the pivot points making the diffraction pattern shift when scanning the beam. Therefore, we opted for a standard off-axis setup where we used a conventional round aperture giving 0.5 mrad convergence angle and we shifted the central disc slightly away from the spectrometer entrance aperture. The results are presented in fig.6.c and show a band gap onset signal that is now independent of the grain orientation and with good spatial resolution over the whole extension of the NCD layer.

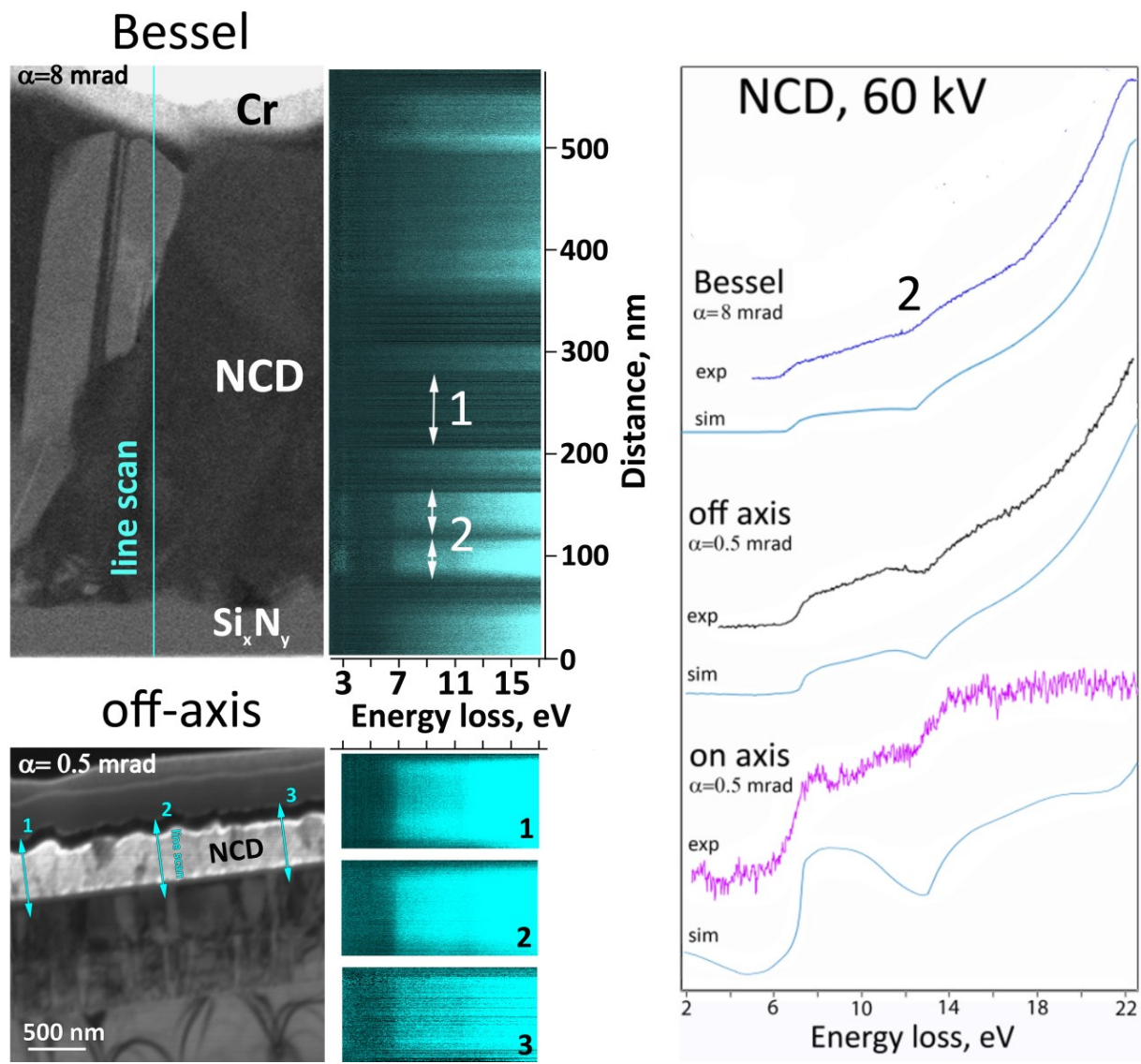

*Fig.6. a. EELS line profile over the NCD layer taken at 60 kV and with a Bessel aperture convergence angle of 8 mrad. b. EELS line profiles over the NCD layer acquired with the same acceleration voltage but with a much smaller convergence angle of 0.5 mrad in off-axis conditions with a conventional round aperture. c. Diamond spectra extracted from the line profiles with different experimental conditions accompanied with the corresponding simulated spectra retrieved from the Kröger equation. In case of off-axis Bessel set-up the band gap signal was extracted from the grains marked with 2. Note that at 8 mrad convergence angle not all the grains demonstrate clear band gap onsets due to diffraction effects and/or collected q-transitions selected too far from the direct band gap. The band gap onsets acquired using small convergence angles are much more consistent over the whole thickness of the polycrystalline diamond layer.*

5. **Conclusions**

The data shown in this article demonstrates the efficiency of off-axis acquisition with Bessel aperture even at high acceleration voltages, as 300 keV, and for materials with high refractive index, like diamond.

Following simple rules such as using relevant convergence angles and preventing the appearance of near-field effects, this technique can be performed at any microscope with a reasonable spectral resolution by simply exchanging a circular condenser aperture for a specifically designed ring aperture.

The interpretation in terms of a simple two band model gives insight in the momentum transfer selection and helps to choose the right acquisition parameters.

**Acknowledgements**

S.K., B.P. and J.V. acknowledge funding from the "Geconcentreerde Onderzoekacties" (GOA) project "Solarpaint" of the University of Antwerp. S.K. and J.V. also acknowledge the FWO-Vlaanderen for financial support under contract G.0044.13N 'Charge ordering'. Financial support via the Methusalem "NANO" network is acknowledged.